\documentclass[twocolumn,showpacs,aps,prb,floatfix]{revtex4}

\usepackage{graphicx}
\usepackage{dcolumn}
\usepackage{bm}

\begin{document}

\preprint{APS/123-QED}

\title{Observation of the Ettingshausen effect in quantum Hall systems}

\author{Yosuke Komori}
 \email{komori@dolphin.phys.s.u-tokyo.ac.jp}
\author{Tohru Okamoto}
 \email{okamoto@phys.s.u-tokyo.ac.jp}
\affiliation{
Department of Physics, University of Tokyo, Hongo 7-3-1, Bunkyo-ku, Tokyo 113-0033, Japan
}%

\date{\today}

\begin{abstract}
Evidence of the Ettingshausen effect
in the breakdown regime of the integer quantum Hall effect
has been observed in a GaAs/AlGaAs two-dimensional electron system.
Resistance of micro Hall bars 
attached to both edges of a current channel
shows remarkable asymmetric behaviors which indicate
an electron temperature difference between the edges.
The sign of the difference depends on 
the direction of the electric current and
the polarity of the magnetic field.
The results are consistent with the recent theory of Akera.
\end{abstract}

\maketitle

A two-dimensional electron system (2DES) at low temperatures and 
in a strong magnetic field shows the quantum Hall effect (QHE),
in which the longitudinal resistivity $\rho_{xx}$
vanishes and the Hall resistance $R_{\rm H}$
in a Hall bar sample is quantized as $R_{\rm H}=h/ie^2$ with an integer $i$.\cite{Kawaji1980,Klitzing1980}
The quantized Hall resistance has been used as an international resistance standard since 1990.\cite{ResSt}
In order to carry out high precision measurements of the quantized Hall resistance,
we must pass a large electric current through a Hall bar to generate
a high Hall voltage across the sample.
When the current exceeds a critical value, however, 
the QHE breaks down with a sharp increase in $\rho_{xx}$ from zero.\cite{Nachtwei1999}
In the breakdown regime, the electron temperature $T_e$ is expected to be 
much higher than the lattice temperature $T_{\rm L}$
due to the Joule heating of electrons.
The hot electrons moving in the electric field 
carry the heat.
In order to understand the mechanism of the breakdown of the QHE, 
it is important to know the spatial distribution of 
$T_e$ and the heat flow in a Hall bar sample.
In the direction of the electric current,
the variation of $T_e$ can be investigated
by measuring the longitudinal resistance 
with a set of voltage probes distributed along the current channel.\cite{Komiyama1996,Kaya1998,Kaya1999,Morita2002}
On the other hand, it has not been studied experimentally
in the direction across the current channel.

In general, the appearance of a temperature gradient or 
a heat flow 
perpendicular to both the electric current and the magnetic field
is known as the Ettingshausen effect.
According to the Onsager reciprocity relations,
it is the cross effect of the Nernst effect,
which has been observed in 2DES's.\cite{Gallagher1992,Fletcher1999}
Very recently, Akera has derived hydrodynamic equations
for quantum Hall systems in the regime of large energy dissipation 
and exhibited the spatial variation of $T_e$
in the direction perpendicular to the electric current.\cite{Akera2002}
His calculation shows that,
although $T_e$ is approximately uniform in the middle of the current channel,
it is higher around one edge and lower around the other.

In this work, we have investigated the Ettingshausen effect
in a two-dimensional GaAs/AlGaAs channel
using micro Hall bars attached to both edges as indicators of $T_e$.
Asymmetric behaviors which correspond to an electron temperature difference
between the edges are observed in the QHE breakdown regime.
The sign of the difference is consistent with Akera's theory.
The thermal relaxation length of hot electrons is also discussed.

The samples are fabricated
on a GaAs/Al$_{0.3}$Ga$_{0.7}$As heterostructure wafer
with an electron density $N_s=2.3 \times 10^{15} {\rm m}^{-2}$
and a mobility $\mu = 36 {\rm m}^2 /{\rm V s}$ at 4.2~K.
As shown in Fig. 1(a),
$600~\mu {\rm m}$ wide current electrodes are
separated from the central part by $1200~\mu {\rm m}$
in order to avoid the effect of electron heating
at the diagonally opposite corners.\cite{comment}
\begin{figure}[b]
\includegraphics{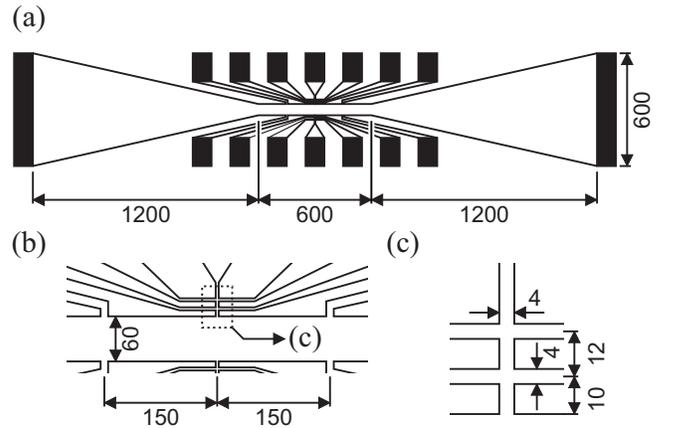}
\caption{
Schematic view of the sample geometry. The numbers are in units of $\mu {\rm m}$.
(a) Overall structure.
(b) Structure of the central part.
(c) Micro Hall bar attached to the main channel.
}
\end{figure}
A similar design was used by Kawaji {\it et al.},\cite{Kawaji1994}
who studied the intrinsic breakdown of the QHE and
found the linear relationship between 
the critical current
and the width of the central channel.
The central part shown in Fig. 1(b) consists of the main channel 
whose width is $60~\mu {\rm m}$,
four voltage probes for the measurements of 
$I$-$V$ characteristics of the main channel,
and two micro Hall bars (MHBs) attached
to the main channel (see Fig. 1(c)).
The longitudinal resistivity in the MHBs
is expected to vary with $T_e$
in the vicinity of each edge.
All the measurements were performed at $T_{\rm L} = 4.2~{\rm K}$ in liquid helium.

Fig. 2 shows the longitudinal voltage drop $V_{\rm MAIN}$ along the main channel
as a function of the electric current $I_{\rm MAIN}$.
\begin{figure}
\includegraphics{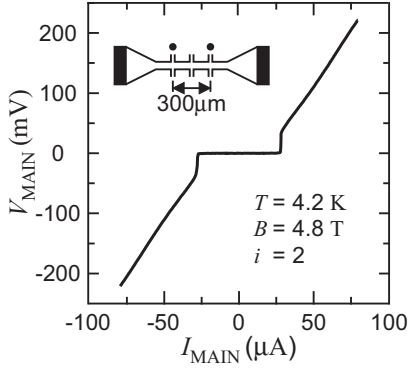}
\caption{
$I$-$V$ characteristics of the main channel.
}
\end{figure}
The magnetic field $B = 4.8~{\rm T}$ corresponds to
the center of the quantum Hall plateau with $i = 2$.
$V_{\rm MAIN}$ abruptly increases at the critical current of
$I_{\rm cr} = 28.6~\mu {\rm A}$,
which is independent of the current polarity
and the voltage probe pair.
The critical Hall electric field $E_{\rm cr}=6.15~{\rm kV/m}$
calculated from $I_{\rm cr}$
is consistent with the previous measurements
for the intrinsic breakdown of the QHE.\cite{Kawaji1994,Okuno1995,Kawaji1996}

A small DC electric current $I_{\rm MHB}$ is used
to draw a part of hot electrons from the main channel
into the MHB.
The differential longitudinal resistance 
$dV_{\rm MHB}/dI_{\rm MHB}$ in the MHB is measured
with an AC modulation current of $0.05~\mu {\rm A}$ and 26.6~Hz
added to $I_{\rm MHB}$.
In Fig. 3, the results with $I_{\rm MHB}= 0.2~\mu {\rm A}$
for different voltage probe pairs
are shown as a function of $I_{\rm MAIN}$.
\begin{figure}
\includegraphics{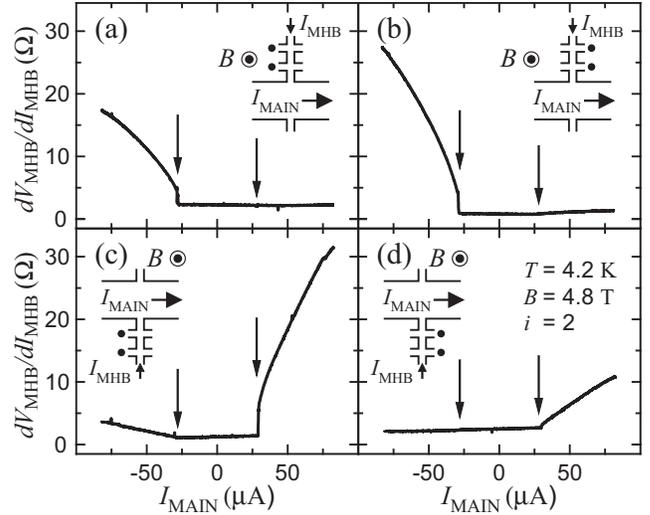}
\caption{
Differential longitudinal resistance $dV_{\rm MHB}/dI_{\rm MHB}$
with $I_{\rm MHB} = 0.2~\mu {\rm A}$
in the upper ((a) and (b)) and lower ((c) and (d)) 
micro Hall bars
as a function of electric current $I_{\rm MAIN}$ through the main channel.
The critical current obtained from Fig.~2 is indicated by arrows.
The black dots in the insets indicate the voltage probes used
in the measurements.
}
\end{figure}
The critical current $I_{\rm cr} = 28.6~\mu {\rm A}$
is indicated by arrows.
The differential longitudinal resistance
in the lower MHB (see Fig. 3(c) and (d))
steeply increases with $I_{\rm MAIN}$ for $I_{\rm MAIN} > I_{\rm cr}$,
while the increase is small
in the upper MHB in the region of $I_{\rm MAIN}>0$ (see Fig. 3(a) and (b)).
This indicates that $T_e$ is higher in the vicinity of the lower edge
than in that of the upper edge. 
The electric potential due to the Hall effect is higher at the upper edge
for $I_{\rm MAIN} > 0$ and the magnetic field direction used here.
The results demonstrate that $T_e$ increases
in the direction of the Hall electric field.
In the case of $I_{\rm MAIN} < 0$, the Hall electric field is reversed
and a steep increase in $dV_{\rm MHB}/dI_{\rm MHB}$
 for $I_{\rm MAIN} < -I_{\rm cr}$ is observed in the upper MHB.
The reversal of the magnetic field also changes the polarity 
as shown in Fig. 4(a).
\begin{figure}
\includegraphics{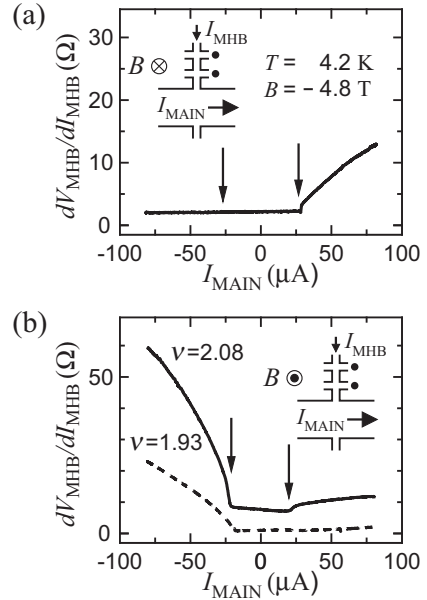}
\caption{
Differential longitudinal resistance $dV_{\rm MHB}/dI_{\rm MHB}$
with $I_{\rm MHB} = 0.2~\mu {\rm A}$
in the upper micro Hall bar.
(a) Data at the reversed magnetic field.
(b) Data at $\nu=1.93$ (dashed curve) with $I_{\rm cr}=19.9~\mu {\rm A}$
and $\nu=2.08$ (solid curve) with $I_{\rm cr}=21.0~\mu {\rm A}$
(indicated by arrows).
}
\end{figure}
Similar asymmetric dependence of $dV_{\rm MHB}/dI_{\rm MHB}$ 
on $I_{\rm MAIN}$ at the center of the QHE plateau
is observed at a lower $T_{\rm L}$ of $1.5~{\rm K}$, 
at a higher $N_s$ of $2.6 \times 10^{15}~{\rm m}^{-2}$ after a LED illumination, 
and in samples fabricated on other wafers.
Fig. 4(b) shows the results at Landau level filling factors
higher or lower than $\nu = 2$ at the center of the QHE plateau.
Overall behaviors are similar to that at $\nu = 2$
while $I_{\rm cr}$ is smaller
and $dV_{\rm MHB}/dI_{\rm MHB}$ for $|I_{\rm MAIN}| < I_{\rm cr}$ is larger.

The temperature difference between both edges indicates
the presence of the component of the heat flow 
perpendicular to the electric current.
Because of $\rho_{xx}>0$ in the breakdown regime,
it is expected that equipotential lines are inclined
 with respect to the channel direction ($x$-direction) as shown in Fig. 5(a). 
\begin{figure}
\includegraphics{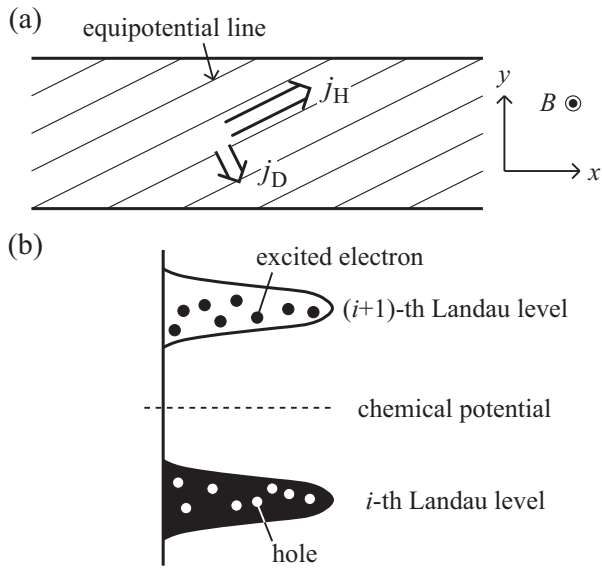}
\caption{
(a) Schematic view of the Hall current and the dissipative current
in a long channel.
(b) Excited electrons and holes in the Landau levels.
The chemical potential lies at the middle of the gap
between the Landau levels at $\nu= {\rm integer}$.
}
\end{figure}
The Hall current flows along the equipotential lines and 
the dissipative current flows across them.
In a sufficiently long channel, the $y$-components
of the Hall current density ${\bm j}_{\rm H}$ 
and the dissipative current density ${\bm j}_{\rm D}$ cancel each other out,
i.e. $j_{\rm H}^y + j_{\rm D}^y =0$.
On the other hand, the sum of the $y$-components
$q_{\rm H}^y$ and $q_{\rm D}^y$
of the heat flow densities 
${\bm q}_{\rm H} $ and ${\bm q}_{\rm D}$ 
is not zero unless the Peltier coefficient
$q_i / j_i$ ($i=$ H or D) is the same for 
both the Hall and dissipative currents.
It is expected that
the $y$-component of the total heat flow 
decreases $T_e$ around one edge
and increases $T_e$ around the other.

The experimental results demonstrate that
the $y$-component of the heat flow
is negative in the case of Fig. 5(a).
The polarity of the Ettingshausen effect 
can be explained by Akera's theory.\cite{Akera2002,AkeraPrv}
Since the direction of the drift motion of electrons
is opposite to the Hall current, the $y$-component $q_{\rm H}^y$
of the heat flow density is negative.
On the other hand,
the dissipative current is carried by 
excited electrons in the higher Landau level 
and holes in the lower Landau level (see Fig. 5(b)),
which move in opposite directions.
At the integer filling factor,
the numbers of excited electrons and holes 
are the same and $q_{\rm D}^y$ cancels out.
Thus the total $y$-component of the heat flow density 
at the center of the quantum Hall plateau is given 
by $q_{\rm H}^y$, which is negative.

As shown in Fig. 4(b),
the polarity of the Ettingshausen effect for $| I_{\rm MAIN} | > I_{\rm cr}$
is the same at $\nu = 1.93$ and 2.08
although $q_{\rm D}^y$ is expected to change its sign
at $\nu = 2$.\cite{Akera2004}
This suggests that $| q_{\rm H}/j_{\rm H} |$
exceeds $| q_{\rm D}/j_{\rm D} |$
in the breakdown regime.\cite{AkeraPrv}
The Peltier coefficient $| q_{\rm H}/j_{\rm H}|$ 
is expected to be large at high $T_e$ since excited electrons and holes
carry the heat toward the same direction along the equipotential lines.
On the other hand, the creation of pairs of 
excited electrons and holes does not contribute to $q_{\rm D}$
while it increases $| j_{\rm D} |$.
In the region of $| I_{\rm MAIN} | < I_{\rm cr}$, 
we  observed a small linear dependence of 
$dV_{\rm MHB}/dI_{\rm MHB}$ on $I_{\rm MAIN}$
for Landau level filling factors
higher or lower than $\nu=2$.
However, the sign of the dependence is negative (positive) for 
$\nu > 2$ ($\nu < 2$)
and opposite to that expected from $q_{\rm D}^y$.
We consider that it arises from a small change in 
the Landau level filling factor in the MHB, 
not from a change in $T_e$.
The capacitive coupling 
between the upper and lower edges
may lead to a small decrease in the local electron density ($\propto \nu$)
in the upper MHB with increasing Hall voltage 
($\propto I_{\rm MAIN}$).
This effect produces a change in the longitudinal resistivity
which steeply varies with $\nu$
except around the center of the quantum Hall plateau.

In order to investigate the thermal relaxation of hot electrons
injected into MHBs,
another sample with a MHB having eight voltage probes
(Fig. 6(a)) was fabricated on the same wafer,
while $N_s$ slightly changes.
\begin{figure}
\includegraphics{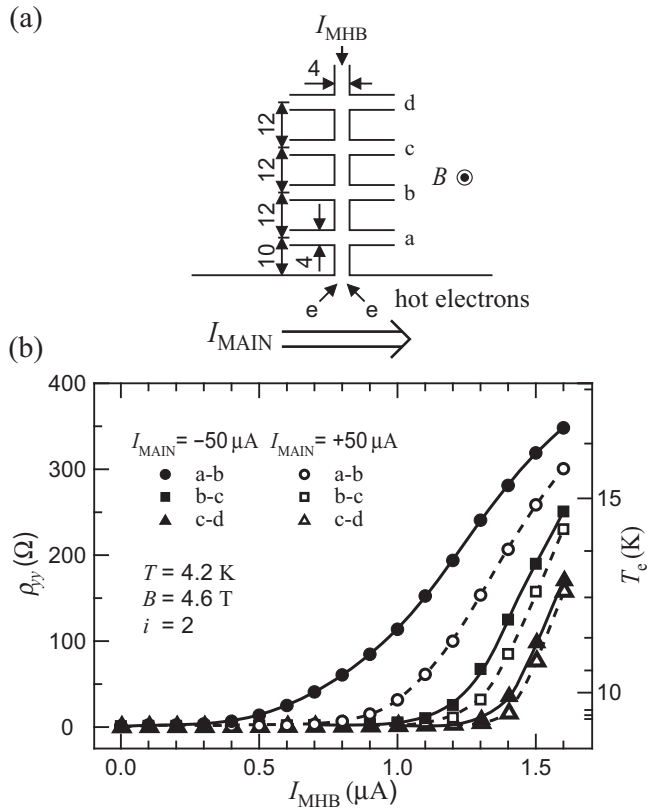}
\caption{
(a) Geometry of the micro Hall bar for thermal relaxation measurements.
The numbers are in units of $\mu {\rm m}$.
(b) Longitudinal resistivity as a function of the electric current.
The full symbols for $I_{\rm MAIN} = - 50~\mu {\rm A}$
and the open symbols for $I_{\rm MAIN} = + 50~\mu {\rm A}$.
Results for various voltage probe pairs are shown.
All the lines are guides to the eyes.}
\end{figure}
The geometry of the other part of the sample is the same as that shown in Fig.~1.
Fig. 6(b) shows the longitudinal resistivity $\rho_{yy}$ in the MHB
for different voltage probe pairs
and different polarities of $I_{\rm MAIN} \pm 50~\mu {\rm A}$.
The magnetic field was fixed at the center of the quantum Hall
plateau with $i=2$, where $I_{\rm cr}=26.9~\mu {\rm A}$.
$T_e$ indicated on the right axis is estimated from
$T_{\rm L}$-dependence of the longitudinal resistivity calibrated  
using a sufficiently small current.
The $I_{\rm MAIN}$-dependence of $\rho_{yy}$ 
shows that the electron temperature
increases with the flux of hot electrons from the main channel.
The dependence of $\rho_{yy}$ on the distance
from the main channel is negative and 
associated with the thermal relaxation of hot electrons
during transport.
In the case of $I_{\rm MAIN}=0$ or $I_{\rm MHB} < 0$ (not shown in the figure),
such dependence on the distance is not observed and 
the increase in $\rho_{yy}$ is very small 
except in the region of
 $| I_{\rm MHB} | \gtrsim 1.5~\mu {\rm A}$
where the intrinsic (local) breakdown occurs in the MHB.
Since $I_{\rm MAIN}$ is not directly related to
the thermal relaxation process in the MHB,
the dependence of $\rho_{yy}$ 
on the polarity of $I_{\rm MAIN}$
should be attributed to the difference in $T_e$
in the vicinity of the upper edge of the main channel
due to the Ettingshausen effect discussed above.

Although the origin of the breakdown of the QHE
is not clear at the present stage,\cite{Nachtwei1999}
the calculation of the critical electric field $E_{\rm cr}$ based on
the heat instability model
gives an upper limit of the relaxation time $\tau_e$
of hot electrons.
Using $E_{\rm cr}=(2 \hbar/m^\ast \tau_e)^{1/2} B$
derived in Ref. 18,
$\tau_e=2.2~{\rm ns}$ is obtained from 
the experimental value of
$E_{\rm cr}=5.79~{\rm kV~/~m}$.
Here, $m^\ast$ is the effective mass of electrons in GaAs
(0.067 of the free electron mass).
If every hot electron in the MHB propagates 
with a mean velocity $v_{\rm MHB}$,
the relaxation length is estimated to be only $1.4~\mu{\rm m}$
or less for $I_{\rm MHB} = 1~\mu {\rm A}$
($v_{\rm MHB}=695~{\rm m/s}$).
This seems consistent with the fact that, 
even for the closest voltage probe pair (a-b) with 
a distance of $10~\mu {\rm m}$ from the main channel,
$\rho_{yy}$ is much smaller than 
the longitudinal resistivity ($\rho_{xx}=610~\Omega$)
in the main channel.
The thermal relaxation length $\ell_{\rm edge}$ that determines
the spatial variation of $T_e$
in the vicinity of the edges
of the main channel can also be estimated from $\tau_e$.
For example, the $y$-component of the drift velocity
along the equipotential lines
is calculated to be $109~{\rm m/s}$
for $|I_{\rm MAIN}|= 50~\mu {\rm A}$.
It leads to a very small value of 
$\ell_{\rm edge}=0.24~\mu {\rm m}$.
The change in $T_e$ due to 
the $y$-component of the heat flow
is expected to be confined to narrow
areas in the vicinity of the edges,
while a random potentials or electron-electron interactions
might change $\ell_{\rm edge}$.

In summary, we have observed evidence of 
the Ettingshausen effect
in the breakdown regime of the integer quantum Hall effect.
The electron temperature difference between the edges of 
a long current channel is explained by
a heat flow along the equipotential lines.

We would like to thank Professor H. Akera and Dr. Y. Kawano
 for valuable discussions. 
This work is supported in part by Grants-in-Aid for Scientific Research 
from the Ministry of Education, Science, Sports and Culture, Japan.


\begin{thebibliography}{30}

\bibitem{Kawaji1980}
S. Kawaji and J. Wakabayashi, in {\it Physics in High Magnetic Fields},
proceedings of the Oji International Seminar, Hakone, Japan, 1980,
edited by S. Chikazumi and N. Miura (Springer-Verlag, Berlin, 1981), p. 284.

\bibitem{Klitzing1980}
K. von Klitzing, G. Dorda, and M. Pepper, Phys. Rev. Lett. {\bf 45}, 494 (1980).

\bibitem{ResSt}
B. Jeckelmann and B. Jeanneret, Rep. Prog. Phys. {\bf 64}, 1603 (2001).

\bibitem{Nachtwei1999}
For a review, see G. Nachtwei, Physica E {\bf4}, 79 (1999).

\bibitem{Komiyama1996}
S. Komiyama, Y. Kawaguchi, T. Osada, and Y. Shiraki, Phys. Rev. Lett. {\bf 77}, 558 (1996).

\bibitem{Kaya1998}
I. I. Kaya, G. Nachtwei, K. von Klitzing, and K. Eberl, 
Phys. Rev. B {\bf 58}, R7536 (1998).

\bibitem{Kaya1999}
I. I. Kaya, G. Nachtwei, K. von Klitzing, and K. Eberl, 
Europhys. Lett. {\bf 46}, 62 (1999).

\bibitem{Morita2002}
K. Morita, S. Nomura, H. Tanaka, H. Kawashima, and S. Kawaji,
Physica E {\bf 12}, 169 (2002).

\bibitem{Gallagher1992}
For a review, see G. L. Gallagher and P. N. Butcher, in {\it Handbook on Semiconductors Completely Revised Edition} edited by T. S. Moss and P. T. Landsberg (Elsevier, Amsterdam, 1992), Vol. 1, Chap. 14, p. 721.

\bibitem{Fletcher1999}
For a recent review, see R. Fletcher, Semicond. Sci. Technol. {\bf 14}, R1 (1999).

\bibitem{Akera2002}
H. Akera, J. Phys. Soc. Jpn. {\bf 71}, 228 (2002).

\bibitem{comment}
In previous works, an asymmetry of the dissipation between the source electrode and the drain electrode has been reported
(U. Kla\ss, W. Dietsche, K. von Klitzing, and K. Ploog, Z. Phys. B {\bf 82}, 351 (1991);
Y. Kawano and S. Komiyama, Phys. Rev. B {\bf 68}, 085328 (2003)).
It is caused by the difference between the entry process of an electron at the source contact and the exit process at the drain contact, and thus not related to the Ettingshausen effect.

\bibitem{Kawaji1994}
S. Kawaji, K. Hirakawa, M. Nagata, T. Okamoto, T. Fukase, and T. Gotoh, J. Phys. Soc. Jpn. {\bf 63}, 2303 (1994).

\bibitem{Okuno1995}
T. Okuno, S. Kawaji, T. Ohrui, T. Okamoto, Y. Kurata, and J. Sakai, J. Phys. Soc. Jpn. {\bf 64}, 1881 (1995).

\bibitem{Kawaji1996}
S. Kawaji, Semicond. Sci. Technol. {\bf 11}, 1546 (1996).

\bibitem{AkeraPrv}
H. Akera, private communication.

\bibitem{Akera2004}
The sign of $q_{\rm D}^y$ is expected to be positive (negative)
when the chemical potential lies above (below) the middle of the gap
between the Landau levels and excited electrons (holes) are
the majority carriers
(H. Akera and H. Suzuura, unpublished).

\bibitem{Komiyama2000}
S. Komiyama and Y. Kawaguchi, Phys. Rev. B {\bf 61}, 2014 (1999).

\end{thebibliography}
\end{document}